\documentclass[12pt]{iopart}

\newcommand{\mymatrix}[1]{
\left( \begin{array}{cc} #1 \end{array} \right) }
\newcommand{\myvector}[1]{
\left( \begin{array}{c} #1 \end{array} \right) }
\renewcommand{\vec}[1]{{\bf #1}}

\usepackage{graphicx}

\begin{document}
\title{Deterministic reordering of  $^{40}${Ca}$^+$ ions in a linear segmented {P}aul trap.}

\author{F.~Splatt$^1$ M.~Harlander$^{1}$ M.~Brownnutt$^1$ F.~Z\"{a}hringer$^{1,2}$ R.~Blatt$^{1,2}$ W.~H\"{a}nsel$^1$ }

\address{$^1$ Institut f\"{u}r Experimentalphysik,
Universit\"{a}t Innsbruck,
Technikerstrasse 25,
A-6020 Innsbruck, Austria}

\address{$^2$ Institut f\"{u}r Quantenoptik und Quanteninformation
der \"{O}sterreichischen Akademie der Wissenschaften,
Technikerstrasse 21a,
A-6020 Innsbruck, Austria}

\ead{max.harlander@uibk.ac.at}

\begin{abstract}
In the endeavour to scale up the number of qubits in an ion-based
quantum computer several groups have started to develop miniaturised
ion traps for extended spatial control and manipulation of the ions.
Shuttling and separation of ion strings have been the foremost
issues in linear-trap arrangements and some prototypes of junctions
have been demonstrated for the extension of ion motion to two
dimensions. While junctions require complex trap structures, small
extensions to the one-dimensional motion can be accomplished in
simple linear trap arrangements. Here, control of the extended field
in a planar, linear chip trap is used to shuttle ions in two
dimensions. With this approach, the order of ions in a string is
deterministically reversed. Optimised potentials are theoretically
derived and simulations show that the reordering can be carried out
adiabatically. The control over individual ion positions in a linear
trap presents a new tool for ion-trap quantum computing. The method
is also expected to work with mixed crystals of different ion
species and as such could have applications for sympathetic cooling
of an ion string.

\end{abstract}

\vspace{2pc}
\submitto{\NJP}
\maketitle
\renewcommand{\thefootnote}{\arabic{footnote}}
\section{Introduction}
\label{sec:Introduction} Ion traps hold great promise for realising
scalable quantum computing and, in most regards, stand as the
pre-eminent system for such applications \cite{ARDA:2004}. The
essential building blocks required for quantum computing
\cite{DiVincenzo:00} have been realised using linear strings of a
few ions \cite{Blatt:08, Haeffner:08}. To fully exploit the power of
trapped-ion quantum computing, such methods must be scaled to
involve many ions which can be made to interact with one another in
different combinations according to the requirements of any
particular algorithm \cite{Steane:07}. One way of achieving this is
to divide the traps into segments, so the ions can be moved and
sorted into arbitrary arrangements \cite{Kielpinski:02}. In such a
system the ability to reorder ions within a single linear ion string
is a highly valuable tool: an ion that carries quantum information
may be passed from one side of an ion string to the other. While it
may seem that such information may instead be passed through the
quantum channel provided by the common ion motion \cite{Cirac:95},
the mechanical method has the advantage that it neither depends on
the electronic encoding scheme, nor relies on precise knowledge of
the trap parameters. This renders the mechanical method very
general. Additionally, ions of different species can be
interchanged. This feature may be of use for sympathetic cooling
applications, where the cooling ions need to be correctly placed
within a longer ion string. While there is a method of sorting ions
of different masses within an ion string through simple Doppler
cooling \cite{Jost:09} this scheme does not apply to ions of similar
masses, \emph{e.g.} $^{40}$Ca$^+$ and $^{43}$Ca$^+$, as recently
used for sympathetic cooling \cite{Home:08}. Additionally, it cannot
achieve arbitrary arrangements of ions, such as an arrangement with
all ions used for sympathetic cooling situated at the same end of
the string, as is required with, \emph{e.g.} ion-trap implantation
techniques \cite{Schnitzler:09}.

In order to interchange the positions of two ions, the ions have to
be moved in (at least) two dimensions. Such 2D-movement can be
achieved in trap architectures that have been specifically designed
for this purpose such as, \emph{e.g.} T-shaped
\cite{Hensinger:06,Hucul:08} or X-shaped \cite{Blakestad:09}
junctions of linear traps.
The reordering of two ions can be
achieved by separating a pair of ions and shuttling them through the junction.
Such junction structures are, however,
technically demanding to fabricate. Additionally, the potential near
the junction is rather complicated and highly sensitive to
(mis)alignment during trap construction.
This can lead to the requirement for careful calibration of the voltages
for each individual path \cite{Hensinger:06, Blakestad:09}.
In \cite{Hensinger:06}, using a T-shaped junction comprising of three layers of electrodes an
exchange success rate of
24\% is reported, together with a heating rate of $\sim$1\,eV per turn.
The heating rate is caused by an RF barrier that---due to the large size of
control electrodes in the junction region---can
only be overcome in a non-adiabatic fashion.
Using a fabrication process which allows narrower
electrodes, the group of D.~Wineland
has demonstrated that
single ions can be reliably shuttled through an X-shaped junction
generated by a two-layer electrode design \cite{Blakestad:09}. Using
fast transport and low-noise RF sources the heating rate was
brought down to $\leq$~15~quanta. Nonetheless, the method still
requires complex trap structures and
rather involved electrode-voltage control to ensure
low heating rates over the RF barrier.
The ability to deterministically
arrange ions within a string using a linear trap would provide a valuable new tool for
segmented ion-trap technology, significantly simplifying the hardware requirements.

In this paper two methods of deterministically reordering ions
within a linear, segmented, planar ion trap are presented. The trap
geometry used \cite{Leibrandt:07} consists of only 5 segments
(\emph{i.e.} RF plus 11 independent DC electrodes) and has no
junction structures. In both methods the ion string is reversed by
an adiabatic change of the trapping potential. In the first method
the trapping potential is rotated about its centre point by
appropriately tailoring the electrode voltages. The technique of
rotating an ion string about its centre is attractive for a number
of practical reasons. Specifically, the optimisation of the
electrode voltages requires the electric fields to be modelled (or
measured) only at a single point. It is also simple to image the
ions during the entire turn, to aid experimental optimisation of the
control voltages. For other realisations of ion reordering
\cite{Hensinger:06,Blakestad:09} the fields must be considered over
an extended path, and imaging the ions at all points of the turn is
more involved or even impossible. The second method demonstrated in
this paper is akin to a ``three-point turn". Here the orientation of
the principal axis with weakest curvature varies as a function of
space, and the ion string is reoriented by exploring this spatially
varying field. The position is controlled only by time-dependent
homogeneous fields. Were these applied to an array of traps it might
be possible to parallelise reordering of ions in multiple traps.

\section{Apparatus}
\label{sec:Apparatus}

\begin{figure}
\begin{center}
\includegraphics{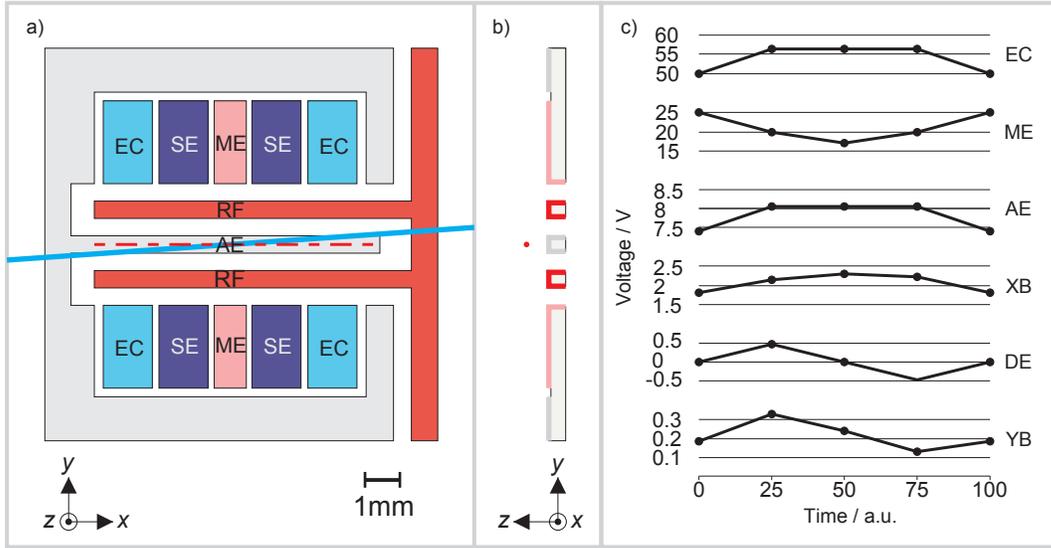}
\parbox{150mm}{\caption
{Schematic drawing of the trap in plan view (a) and cross section
(b), showing RF electrodes (RF), axial electrode (AE), endcap
electrodes (EC), middle electrodes (ME), and steering electrodes
(SE). The red dashed line (a) / dot (b) indicates the position of
the RF null. The blue tilted line in (a) indicates the orientation
of the laser beams.
 (c) Voltage ramps applied to the
specified electrodes for a single exchange, which rotates the ion
string through 180$^{\circ}$. As described in
figure~\ref{fig:BasisSets} the steering electrodes are combined as
$x$-balance (XB), diagonal electrodes (DE), and $y$-balance (YB).
The offset is set to 0\,V.} \label{fig:ElectrodeVoltages}}
\end{center}
\end{figure}

For both reordering methods the ions are held in a linear,
segmented, planar ion trap which consists of 5 segments, as shown in
figure~\ref{fig:ElectrodeVoltages}(a-b). The electrodes are made of
copper on a vacuum-compatible printed-circuit-board substrate, with
an ion-surface distance of around 830~$\mu$m \cite{Leibrandt:07}. A
radially confining potential is produced by an RF voltage of
amplitude $V_{\rm 0}\sim300$~V (0-peak) and frequency $\Omega_{\rm
T}$~=~$2\pi\times$10.125~MHz applied to the RF rails
(figure~\ref{fig:ElectrodeVoltages}~RF). The voltage amplitude is
measured using a capacitive divider just before the vacuum
feedthrough. There is $\sim$10\% uncertainty on the value of the RF
voltage at the trap. Axial confinement is provided by a DC voltage
of $\sim$~55~V applied to the four endcap electrodes
(figure~\ref{fig:ElectrodeVoltages}~EC). DC voltages of up to 25~V
are applied to the remaining 7 electrodes to tailor the potential.
This set-up yields typical motional frequencies around ($\omega_{\rm
x}$,~$\omega_{\rm y}$,~$\omega_{\rm z}$) =~2$\pi\times$(120, 230,
790)~kHz (where the subscripts denote motion in the directions as
defined in figure~\ref{fig:ElectrodeVoltages}). The values of the DC
voltages are controlled by two analog output boards (National
Instruments, NI-PCI6733). These voltages are then filtered by RC
filters with a cut-off frequency of 1~kHz, situated outside the
vacuum chamber, around 25~cm from the trap.

The energy levels of $^{40}$Ca$^+$, and the laser systems required
for cooling and manipulation of the ions, are described in detail
elsewhere \cite{Schmidt-Kaler:03a}, and summarised here. Laser light
at 397~nm is used for Doppler cooling and state detection on the
S$_{1/2}$--P$_{1/2}$ transition, with a repumper laser at 866~nm
tuned to the D$_{3/2}$--P$_{1/2}$ transition. A laser at 729~nm is
used to excite the ion from the ground state, S$_{1/2}$ (hereafter
referred to as $|S\rangle$), to the long-lived D$_{5/2}$ state,
($|D\rangle$). Light at 854~nm can then be used to quench the
$|D\rangle$ state and to rapidly return the ion to the ground state
via the P$_{3/2}$ state. All laser beams are parallel to the surface
of the trap, and at an angle of 4$^{\circ}$ to the $x$-axis, as
defined in figure~\ref{fig:ElectrodeVoltages}(a).

Fluorescence light at 397~nm is imaged using a custom-made lens
(f\#~=~1.7, NA~=~0.28, focal length~67\,mm)
onto a CCD camera (Andor iXon DV885 JCS-VP)
to provide a resolution of 2~$\mu$m per pixel. The ions' micromotion
is coarsely minimised in the $y$-direction by ensuring that the
position of the ions observed on the camera does not vary as a
function of RF power \cite{Berkeland:98}. The height of the ions
above the trap ($z$-direction) is adjusted to ensure that the ions
are well crystalised (\emph{i.e.} each individual ion's image is
well localised on the CCD camera).

\section{One-point turn in a planar trap}
\label{sec:OnePointTurn}

To perform a rotation of the ion string the trap's axial confinement
(along $x$) is steadily increased until it is stronger than the
weakest axis of the radial confinement (along $y$), so that the ion
string is aligned perpendicular to the radio-frequency null axis.
Diagonal pairs of electrodes are used to ensure that---as the axial
confinement is changed---the ion string smoothly aligns itself to
the appropriate direction. When the axial confinement is relaxed,
opposite voltages of the diagonal electrodes cause the ion string to
continue its rotation, and finally to reach its original position
with reversed order.

\begin{figure}
\begin{center}
\includegraphics{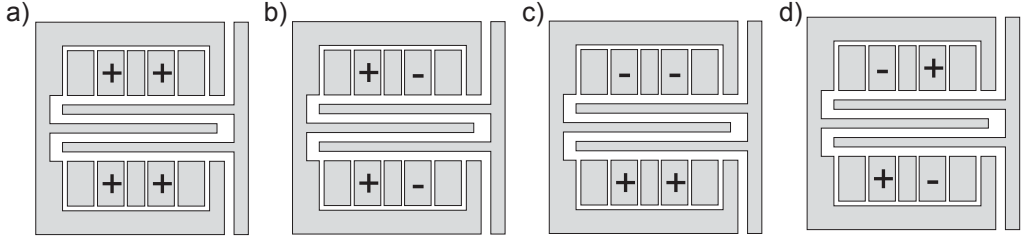}
\parbox{150mm}{\caption
{Four possible ways in which the ``steering electrode" voltages can be varied.
These are termed (a) offset, (b) $x$-balance, (c) $y$-balance, (d) diagonal.}
\label{fig:BasisSets}}
\end{center}
\end{figure}

To easily control and understand the action of the four so-called
``steering electrodes" (figure~\ref{fig:ElectrodeVoltages}~SE),
their voltages are defined in a particular basis as illustrated in
figure~\ref{fig:BasisSets}. The configurations are termed:
\emph{offset}, in which the potential of all four electrodes is
moved together; \emph{x-balance}, in which a potential difference is
applied between the left and right electrode pairs;
\emph{y-balance}, in which a potential difference is applied between
the top and bottom pairs; and \emph{diagonal}, in which a potential
difference is applied between the diagonal pairs. The offset has the
effect of moving the ions along the $z$-direction, and the $x$- and
$y$-balance move the ions in the plane of the trap. The diagonal
configuration creates a potential with a saddle point at the middle
of the trap, oriented with axes of symmetry at 45$^{\circ}$ to the
$x$- and $y$-directions. This allows the principal axes of motion to
be rotated smoothly between $x$ and $y$.

A typical set of voltages used to perform a turn is shown in
figure~\ref{fig:ElectrodeVoltages}(c). A qualitative explanation of
the voltages required is as follows. The very high confinement along
the $z$-axis (perpendicular to the substrate) compared to that along
the two directions parallel to the substrate ($\omega_{\rm
z}/\omega_{\rm y}\sim$~3.5, $\omega_{\rm z}/\omega_{\rm x}\sim$~6.5)
means that, even for reasonably large variations (several volts) in
the DC potential, the confinement in $z$ remains strong compared to
that in both $x$ and $y$. This largely simplifies the problem so
that only the $x$-$y$-plane need be considered. As detailed in
section \ref{sec:TheoryRotation} the crucial part to achieve an ion
rotation is a sufficient compression along the RF null (here, the
$x$-axis) as compared to the weak radial axis (here, the $y$-axis).
For the given trap layout this compression is best reached by a
combination of increasing the endcap electrode voltage
(figure~\ref{fig:ElectrodeVoltages}~EC) and decreasing the middle
electrode voltage (figure~\ref{fig:ElectrodeVoltages}~ME). Between
the two extremes (aligned along $x$, aligned along $y$) the diagonal
electrodes are used to smoothly vary the principal axes of the ions'
motion. Given the asymmetric nature of the trap (\emph{i.e.} that
all electrodes are situated in a single plane below the ion) a
change in the endcap and middle electrode voltages also affects the
height of the ions above the trap surface ($z$-direction). The
axial-electrode voltage (figure~\ref{fig:ElectrodeVoltages}~AE) is
adjusted to ensure that the ion string remains well crystalised.
Finally, to compensate for the effects of any static or patch
charges, or any imperfections in the trap fabrication, the $x$- and
$y$-balance are set to ensure that the ion string stays centred on a
single position during the turn.

\begin{figure}
\begin{center}
\includegraphics{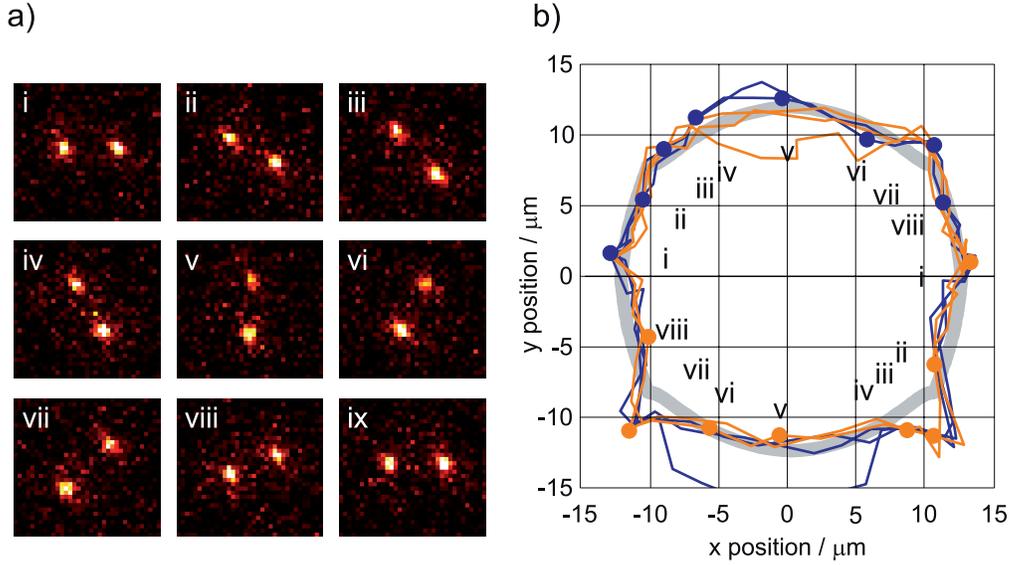}
\parbox{150mm}{\caption
{(a) CCD camera images of a pair of ions at different stages of a
turn. An animation of this motion is given in the online material.
(b) Paths taken by two ions over several turns. The thin lines show
the experimentally measured paths of the two ions, coloured blue and
orange respectively. The dots, labeled by numerals, correspond to
the ion positions shown in (a). The thick grey line indicates the
simulated path. When the ion string is perpendicular to the
Doppler-cooling beam (approximately at position v), it is heated in
the axial direction. This explains the greater variation in path
taken in those areas. For operation without illumination this
heating mechanism is not present.} \label{fig:Path}}
\end{center}
\end{figure}

To create the voltage ramps required for a turn, a string of three
or more ions is observed with the CCD camera, and the electrode
voltages are adjusted in real time to place the ions in certain key
positions (namely, centred on a fixed point, with rotations of
0$^{\circ}$, 45$^{\circ}$, 90$^{\circ}$ and 135$^{\circ}$). In
addition to constraining the centre position and angle of the
string, voltages are chosen to ensure that the string remains in a
linear configuration, rather than adopting a zig-zag form. This sets
a lower bound on the ratio of the transverse to longitudinal
motional frequencies \cite{Schiffer:93}. In the case of a 3-ion
string this is $\omega_{\rm T}/\omega_{\rm L}$~$\geq$~1.6. A linear
interpolation between the electrode voltages for the key positions
is then made, to give a continuously varying ramp, as shown in
figure~\ref{fig:ElectrodeVoltages}(c). Using these voltage ramps the
ion positions were exchanged and a pair of ions traced out the paths
shown in figure~\ref{fig:Path}. The rotation of the ions is also
shown as an animation in the online material. The inter-ion
separation varied between 21 and 31~$\mu$m. Correspondingly, the
longitudinal frequency of the chain varied in the range 75~kHz
$<$~$\omega_{\rm L}$ $<$~125~kHz.

The path measured in the experiment agrees well with
boundary-element method simulations, carried out using
CPO\footnote{Charged Particle Optics Version~7.1. Free version
available at www.electronoptics.com}. For these calculations, the DC
potential was simulated using the exact potentials applied to the DC
electrodes. There is a small uncertainty for the RF drive voltage.
We display the path for an RF voltage of $V_{\rm RF}$~=~300~V, which best
fits the observed ion positions and is, within the error margin,
consistent with the measured value of 320~V. Equally, the
$z$-position, information about which is not accessible through our
camera observation, has been assumed to be constant at
$z$~=~880~$\mu$m, in close proximity to the expected height of the
RF null, $z_{\rm 0}$=830\,$\mu$m. With these values, the calculated ion
positions are very close to the ones observed, as shown in
figure~\ref{fig:Path}(b). Given this ability to model the potentials,
the voltage ramps can in principle be optimised to yield constant
longitudinal and transverse frequency during the rotation.
Theoretical considerations regarding the realisation of such smooth
rotations are given in section~\ref{sec:TheoryRotation}.

\begin{figure}
\begin{center}
\includegraphics{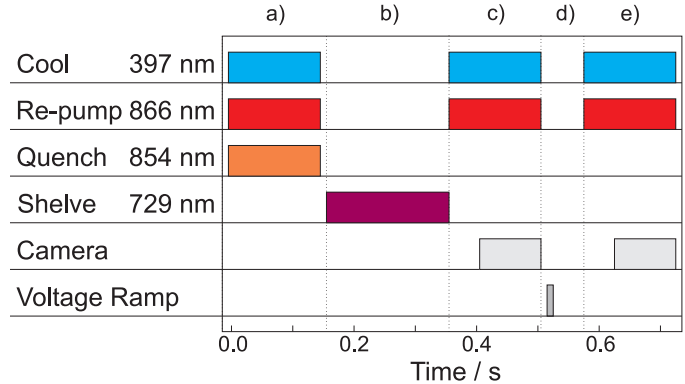}
\parbox{150mm}{\caption
{Pulse sequence for the determination of the exchange fidelity. Ions
are Doppler-cooled (a), labeled by electron shelving (b-c), and
rotated without laser illumination (d). Finally, the resulting ion
position is read out (e).} \label{fig:PulseSequence}}
\end{center}
\end{figure}

To measure the fidelity of the procedure, an exchange of a pair of
Doppler-cooled ions is performed without laser illumination. First
the ions are Doppler-cooled and, by means of the repump laser at
854\,nm, prepared in the $|S\rangle$ state
(figure~\ref{fig:PulseSequence}(a)\,). Next a 729\,nm laser pulse is
commonly addressed to both ions so that each ion is independently
excited to the $|D\rangle$ state with an average probability of
$\approx$20-50\% (b). An additional Doppler-cooling period with
camera detection is then used to select the cases in which exactly
one ion has been shelved (c). The ion chain is then rotated (d) and
again imaged to identify the position of the shelved ion (e). Cases
in which the shelved ion has decayed back to the $|S\rangle$ state
are disregarded as they contain no information.

With this method, the fidelity of exchange was determined to 97\%
(433 events), and within the counting statistics, no significant
dependance on the direction of the turn was observed (clockwise:
189/193, anticlockwise: 232/240). Furthermore, the fidelity showed
no dependence on the exchange time, $T_{\rm swap}$, in the range
1.5\,ms$~<~T_{\rm swap}~<$~20~ms. Below 1.5\,ms the fidelity dropped
rapidly due to the low-pass filters significantly altering the
waveform applied to the electrodes. From numerical simulations,
described in section~\ref{sec:InducedHeating}, we believe that the
exchange could be made a factor of ten times faster if the filters
were adapted in frequency and the reorientation of the trap axis was
performed in a smooth way. For motional frequencies as high as those
used in \cite{Blakestad:09}, the reordering could be performed
within as little as 10\,$\mu$s. It is worth noting that such a time
scale is fast even compared to a conventional coherent swap gate
\cite{Steane:00,Gulde:03}.

If the ions were held for the duration of the exchange routine,
without applying the exchange voltage-ramp sequence, the probability
that they maintained their original ordering was measured to be 97\%
(182 events). The fidelity of the exchange seems therefore to be
presently limited by the rate at which the ions are able to exchange
positions within the standard linear trap. The origin of this
stochastic exchange mechanism has not been fully investigated,
although it may be attributable to the comparatively high background
pressure of $\sim10^{-9}$\,mbar and to imperfect Doppler cooling,
particularly in the radial directions, due to the orientation of the
Doppler-cooling beam.

\section{Induced heating}
\label{sec:InducedHeating} Beyond the efficacy of the reordering, a
figure of merit for the swap operation is the ion-heating induced by
the time-varying potentials. In performing a turn as described, the
ions necessarily leave the RF null by half the inter-ion distance
(here, by 12\,$\mu$m). Carrying out a turn with a greater number of
ions will lead to even larger excursions into regions of significant
RF potential. It may be thought that, while the micromotion is
conservative for a single ion, micromotion could nonetheless couple
to secular motion if there is more than one ion in trap, and cause
heating \cite{Berkeland:98}. However, for crystals with small ion
numbers ($<$5) the micromotion-driven motional spectrum has been
shown to be dominated by a small number of discrete frequencies
and---being essentially integrable---offers no heating mechanism in
the absence of laser light \cite{Bluemel:89,Bluemel:88}. By leaving
the RF null the ions, for the exchange operation parameters
described above, the ion traverses an effective RF barrier of
0.4\,meV. The lack of heating across this barrier is confirmed by
numerical simulations of an ion exchange using optimised electric
fields which preserve the trap frequencies given above (the full
description of the fields is given in
section~\ref{sec:TheoryRotation}).

\begin{figure}
\begin{center}
\includegraphics[width=11.0cm]{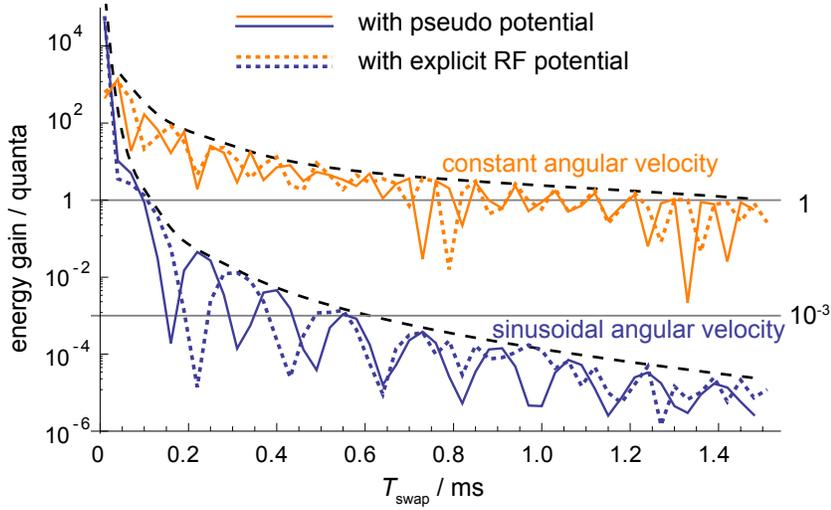}
\parbox{150mm}{\caption
{Numerical simulations of heating as a function of swap duration.
The upper (orange) and lower (blue) curves show the total energy
gain per exchange with constant and variable angular velocity
respectively. The dips in the curves occur when a resonance
condition is met between the vibrational frequencies of the two ions
and the swap rate. The solid and dotted lines give the results of
the calculations with and without the pseudopotential approximation,
respectively. The dashed lines provide a guide to the eye indicating
the envelopes for the curves, and show that inclusion of micromotion
(\textit{i.e.} the use of the explicit RF potential) does not impact
the induced heating rate. With a sinusoidal variation of the
velocity, a swap time of 120\,$\mu$s would lead to an energy gain of
only one vibrational quantum.} \label{fig:Heating}}
\end{center}
\end{figure}

The classical path of two trapped ions has been computed for
different durations of the swap, and the total acquired energy
(kinetic plus potential energy) at the end of the sequence is
calculated. Figure~\ref{fig:Heating} shows the acquired energy in
units of longitudinal vibrational quanta, $\hbar\omega_{\rm L}$. The
two upper curves refer to a swap with constant angular velocity
throughout the turn, \textit{i.e.} with an abrupt start and stop.
The two lower curves result from a rotation with angular-velocity
that varies according to the first half of a sine function. For both
cases, the dynamics have been computed using first the
pseudopotential approximation, and subsequently the full
time-dependence of the RF field. The results show differences in the
detailed structure, while the envelope for the acquired energy is
unchanged by the explicit inclusion of the radio-frequency. At an
exchange duration of $T_{\rm swap}$=1.5\,ms, even for abrupt
velocity changes, a total energy gain equivalent to only one
motional quantum is predicted. For smooth variation of the trap
orientation the same low degree of heating can be reached with a
swap duration of 120\,$\mu$s. This constitutes a near-adiabatic
exchange of the ions.

While no direct heating measurements have been made in our
apparatus, an upper limit on the heating rate induced in the
experiment was inferred by a sequence of 200 consecutive exchanges
performed after previous Doppler cooling. The sequence was repeated
149 times before an ion was lost from the trap. Computer simulations
for the specified operating conditions yield a trap depth of
$\sim$1\,eV. Assuming a thermal distribution of the ions' kinetic
energy after the 200 swap operations, the observed loss rate
suggests a final mean kinetic energy of not more than 150\,meV, at
the 99\% confidence level. Additionally assuming a linear heating
mechanism, this corresponds to an estimated heating rate upper
bounded by $\sim$0.8\,meV per exchange. Even this upper bound
compares favourably with the heating rate of $\sim$1\,eV per
exchange reported when reordering ions through a T-junction
\cite{Hensinger:06}. The significant difference in heating is
because the large electrode extent in \cite{Hensinger:06} precluded
adiabatic shuttling over the RF barrier. The estimated upper bound
in our experiment is still almost four orders of magnitude above the
heating rate measured when moving pairs of ions through an
X-junction \cite{Blakestad:09}, and five orders of magnitude above
the calculated optimal transport protocoll for this trap. For proper
comparison of methods, more precise temperature measurements would
need to be made.

\section{Three-point turn in a planar trap}
\label{sec:ThreePointTurn}

\begin{figure}
\begin{center}
\includegraphics[width=12.0cm]{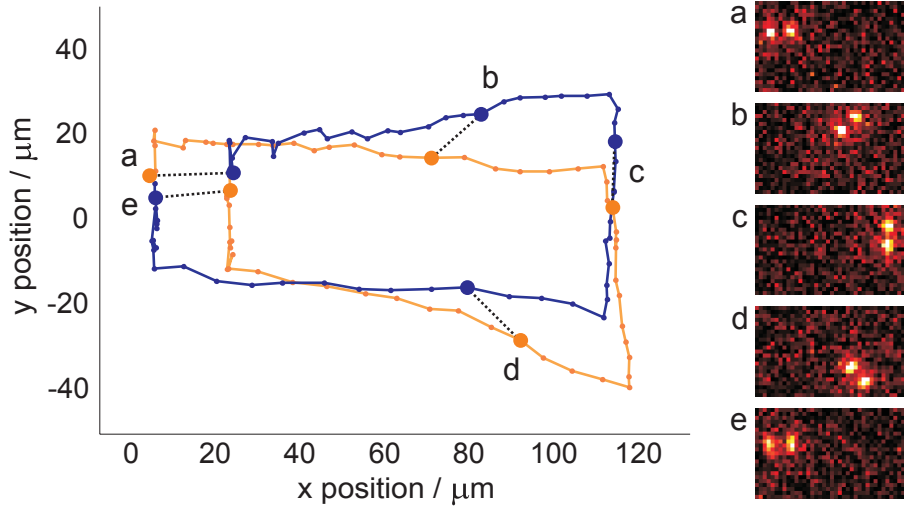}
\parbox{150mm}{\caption
{An ion string can be rotated by using $x$- and $y$-balance to
sample different parts of a potential with spatially varying
anisotropy. By moving off the RF null, the change from axial to
radial realignment is made continuous. The small blue and orange dots
indicate the respective measured positions of the two ions at
different times during the turn (joined by straight lines to guide
the eye). For five example positions marked by larger circles and
joining dotted lines (a-e), CCD images are shown to illustrate the
orientation of the ion pair.} \label{fig:3PTPath}}
\end{center}
\end{figure}

Rather than creating a variable trap anisotropy at a given point by
changing all of the electrode voltages, the electrodes can be set
such that the trap anisotropy varies spatially in an appropriate
fashion. The $x$- and $y$-balance (which generate negligible
quadrupole components over the region of interest, and thus do not
significantly alter the anisotropy themselves) are used to move the
ions to different points in this spatially varying potential.
Typical voltages are shown in table~1. These (excluding $x$- and
$y$-balance) provide the background potential. The path taken by the
ions is shown in figure~\ref{fig:3PTPath}. At the starting position
(a) the ions are aligned along the $x$-axis, while at point
(c)--100\,$\mu$m away--they are aligned along $y$. To move directly
from (a) to (c) would cause the ions to jump randomly to
$\pm$90$^{\circ}$. The reordering can be made deterministic by
passing through the off-axis points (b) and (d) in
figure~\ref{fig:3PTPath}. With this method an exchange fidelity
of~93\% was observed. This value is limited by the relatively small
anisotropy attained near the intermediate points (b) and (d),
\textit{i.e.} at these points the ratio of the longitudinal to the
lower transverse motional frequency approaches unity, so that the
trap orientation is only weakly defined. In principle, the
reliability of exchange can be improved by optimising the path
taken, to keep this anisotropy consistently high. While under the
given conditions this method may not seem preferable to the rotation
about a point on the RF null, it illustrates that ion strings may be
taken well away from the RF null for a short excursion. This adds
flexibility to the net potentials seen by the ions, and may be
particularly interesting at a junction between two linear
radio-frequency traps. The use of spatially homogeneous potentials
may also find applications for parallel control of many operations
in arrays of traps.

\begin{table}[h]
\begin{center}
\begin{tabular}{l l}
\hline \hline
Electrode           & Voltage / V \\
\hline
Endcaps         & 47    \\
Axial electrodes  & 10    \\
Middle electrodes & 30    \\
Offset          & 7.5   \\
$x$-Balance         & 6.5~$\pm$0.7  \\
$y$-Balance         & 0.3~$\pm$0.2  \\
Diagonal electrodes& 0.0\\
\hline \hline
\end{tabular}
\parbox{120mm}{{\vspace{3mm}
\textbf{Table~1.} Typical electrode voltages for a three-point turn.
Electrode names are as defined in
figures~\ref{fig:ElectrodeVoltages} and \ref{fig:BasisSets}.}
\label{tab:3PTVoltages}}
\end{center}
\end{table}

\section{Considerations for a turn in a three-dimensional segmented trap}
\label{sec:TheoryRotation}

In the planar trap that is used in these experiments the ion
movement is easily confined to two dimensions because the
confinement along the $z$-axis (perpendicular to the trap surface)
is by far the strongest, as described above. This is due, in part,
to the ions being situated outside of the plane of the electrodes,
far from the symmetry axis of the trap structure (as seen in
figure~\ref{fig:ElectrodeVoltages}(b)\,). In many other trap
geometries, however, the ions are trapped on the physical symmetry
axis of the electrode structure, \emph{e.g.} in two-layer segmented
traps \cite{Rowe:02, Stick:06, Schulz:08}. It is therefore important
to consider whether the ability to perform the ion string rotation
arises from the fact that the ions are trapped away from the
symmetry axis of the electrode structure, or whether the results
demonstrated here can be generalised to other trap geometries. As a
particular example, the segmented Paul trap dedicated to single-ion
implantation into crystals \cite{Schnitzler:09} requires dopant ions
to be sympathetically cooled by $^{40}$Ca$^+$ ions before being
expelled towards the target. For these experiments it would be a
real advantage if a reordering strategy similar to the one presented
here could be implemented.

\begin{figure}
\begin{center}
\includegraphics[width=0.75\textwidth,keepaspectratio]{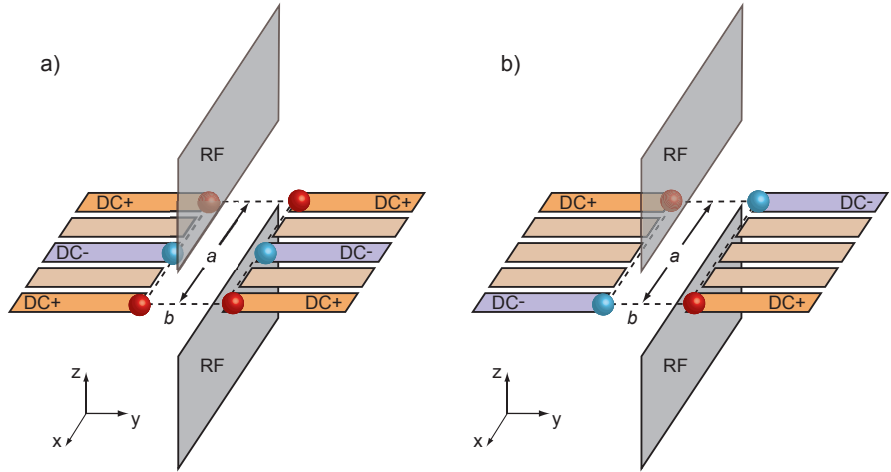}
\parbox{150mm}{\caption
{Model geometry for a three-dimensional segmented ion trap and the
associated point charge model used to optimise the layout of control
electrodes. The figures (a) and (b) illustrate two different voltage
configurations that are used to generate the transverse and the
diagonally oriented quadrupole fields respectively. The resulting
fields are sketched in figure~\ref{fig:Waveforms}. The RF axis has
been chosen along the $x$-axis to be consistent with the convention
used for the planar trap.} \label{fig:TrapModel3D}}
\end{center}
\end{figure}

In the following, the general requirements for the rotation of an
ion string are considered, and appropriate geometries to
generate the required fields in a three-dimensional segmented trap 
are discussed. A generic geometry, as illustrated in
figure~\ref{fig:TrapModel3D}, is assumed (this also directly
translates into standard two-layer segmented traps), and a simple
point-charge model is used to choose the optimal spacings for the
control electrodes. In this geometry the pseudopotential due to the
radio-frequency voltage confines the ions along the $y$- and
$z$-axes, and leaves them free along the $x$-axis. Maxwell's
equation, $\nabla^2 U = 0$ in free space, then requires that the
radio-frequency confinement along $y$ and $z$ is equal. The DC
voltages on the segmented electrodes lead to an additional confining
or repelling potential and, again from Maxwell's equation, it
follows that the net sum of this potential's curvatures along the
three axes is zero.

The DC potentials can be used to continuously rotate the weakest
trap axis so that the ion string smoothly follows the orientation of
the trapping potential. During rotation, a minimum ratio of the
second-lowest to lowest trap frequency must be preserved to prevent
the ion string from folding into a zig-zag configuration
\cite{Schiffer:93}. When the ion string is oriented transverse to
the RF null, the DC confinement along the RF-null axis (here
$\vec{e}_{\rm x}$) has to be strongly increased, while it is
weakened along one radial axis -- in the following labeled $y$-axis.
Moreover, the curvature of the potential along the principal
trapping axes (\emph{i.e.}\ the $z$-axis and the two time-dependent
principal axes in the $x$-$y$-plane) should be kept constant in
order to avoid heating.

\begin{figure}[htb]
\begin{center}
\includegraphics[width=0.7\textwidth,keepaspectratio]{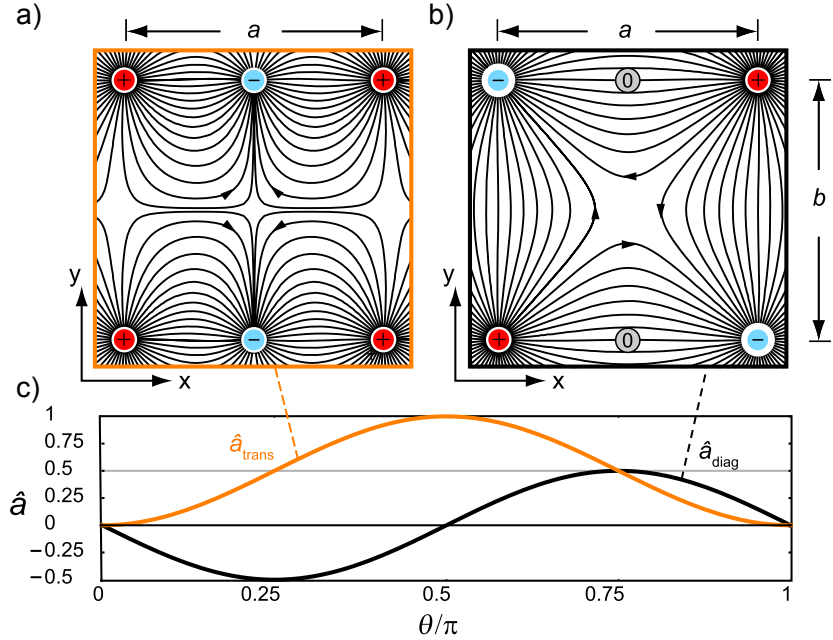}
\parbox{150mm}{\caption
{Two-dimensional quadrupole fields involved in the ion string
rotation. The plots (a) and (b) show the field lines associated with
the transverse and diagonal quadrupole potentials, $\Delta U_{\rm
trans}$ and $\Delta U_{\rm diag}$ respectively. The red and blue
circles indicate the position of the point charges representing the
electrodes (these are the same configurations as in
figure~\ref{fig:TrapModel3D}(a) and (b) respectively. Note that the
RF-null is oriented along the $x$ axes). In configuration (b) no
charges are assumed at the position of the middle electrodes.  The
sinusoidal curves in graph (c) represent the evolution of the
respective fields' strengths during a half rotation as computed in
 equation~(\ref{eq:waveform}).}
\label{fig:Waveforms}}
\end{center}
\end{figure}

Such a rotation of the axes can be achieved by a time-dependent
two-dimensional quadrupole field, $\Delta U(t)$, with its principal
axes in the $x$-$y$-plane. As will be shown below, this field can be
created from the superposition of two separate quadrupole fields, in
the following labeled $\Delta U_{\rm trans}$ and $\Delta U_{\rm
diag}$, with the principal axes oriented along $\vec{e}_{\rm x}$,
$\vec{e}_{\rm y}$ for the first field, and diagonal to these for the
second field. In a typical three-dimensional segmented trap, the
required potentials can be created by appropriate voltages on six DC
electrodes as illustrated in figures~\ref{fig:TrapModel3D} and
\ref{fig:Waveforms}. The field to be added can thus be written as
 \begin{equation}
 \Delta U(t)=\hat a_{\rm trans}(t)\,\Delta U_{\rm trans} + \hat a_{\rm diag}(t)\,\Delta U_{\rm
 diag}\,.
 \end{equation}

To establish this result, and the ideal time evolution of the
relative quadrupole amplitudes $\hat a_{\rm trans}$ and $\hat a_{\rm
diag}$, a linear ion trap is considered, where an initial potential,
$U_{\rm 0}$, confines the ions aligned along the $x$-axis. The desired
potential, $U(t)$, is simply the initial potential, $U_{\rm 0}$, rotated
by an angle, $\theta(t)$, around the $z$-axis. The additional
potential, $\Delta U$, required to achieve this rotation is
determined through the relation
 \begin{equation}
  \Delta U(t)=U(t)-U_{\rm 0}  \,. \label{eq:DeltaU}
 \end{equation}
The initial effective potential, $U_{\rm 0}$, is given by:
\begin{eqnarray}
 U_{\rm 0}&=&\frac12 (\kappa_{\rm DC,x})x^2 +
  \frac12 (\kappa_{\rm RF}+\kappa_{\rm DC},y)y^2
  + \frac12 \left(\kappa_{\rm RF}-(\kappa_{\rm DC,x}+\kappa_{\rm DC,y})\right)z^2  \\[1mm]
  &\equiv &\frac12 \kappa_{\rm x}\,x^2 +
  \frac12 \kappa_{\rm y}\,y^2 +
  \frac12 \kappa_{\rm z}\,z^2 \, ,
 \label{eq:confinement}
 \end{eqnarray}
with, in general, three different curvatures, $\kappa_i$, that are
the sum of curvatures $\kappa_{{\rm DC},i}$ and $\kappa_{{\rm
RF},i}$ arising from the static potential and the RF pseudopotential, respectively.

In order to transform this potential into one where the ion string
is aligned along the $y$-axis, the respective potential curvatures
along the $x$- and $y$-axes need to be swapped. This is exactly
achieved by a two-dimensional quadrupole field oriented along these
axes,
 \begin{equation}
 \Delta U_{\rm trans}=\frac12 \Delta\kappa (x^2-y^2) \,,
 \label{eq:DeltaKappaTransverse}
 \end{equation}
where the curvature $\Delta\kappa= \kappa_{\rm y}-\kappa_{\rm x} $.

This transverse quadrupole field alone is, however, not sufficient
to smoothly rotate the ion trap to its transverse position. The
mirror symmetry about the $x$-$z$-plane needs to be broken in order
to define the direction of rotation. This is achieved by a second
two-dimensional quadrupole field, $\Delta U_{\rm diag}$, of same
strength that has its axes rotated by 45$^{\circ}$ with respect to
$\vec{e}_{\rm x}$ and $\vec{e}_{\rm y}$. This calculation is
facilitated using quadratic forms:
 \begin{equation}
 \Delta U = \frac12 (x,y)\cdot Q \cdot \myvector{x\\y} \,.
 \end{equation}
The two quadrupole fields are then characterised by the two square
matrices
 \begin{eqnarray}
 Q_{\rm trans}& = & \Delta \kappa \mymatrix{1&0\\0&-1} \\
 Q_{\rm diag} & = & R(\pi/4)\cdot Q_{\rm trans}\cdot
 R(-\pi/4)=\Delta\kappa \,\mymatrix{0&1\\1&0} \,,
\end{eqnarray}
where $R(\theta)$ denotes the matrix for an anticlockwise rotation
about the angle $\theta$:
\begin{equation}
 R(\theta)=\mymatrix{\cos(\theta)&-\sin(\theta) \\ \sin(\theta) &
 \cos(\theta)} \,.
\end{equation}
(The diagonal quadrupole potential is then $\Delta U_{\rm
diag}=\Delta \kappa\cdot x\,y$.)

The additional potential, $\Delta U(\theta)$, needed to transform
the initial trapping potential into one rotated by an angle
$\theta$, is given by equation \ref{eq:DeltaU} and is thus expressed
through the square matrix
\begin{equation}
 Q(\theta) = R(\theta)\!\cdot\! \mymatrix{\kappa_{\rm x} & 0 \\ 0 & \kappa_{\rm y}}\!\cdot\! R(-\theta) \,\,-\,\mymatrix{\kappa_{\rm x} & 0 \\ 0 &
 \kappa_{\rm y}}\,,
\end{equation}
which finally equates to
 \begin{equation}
 Q(\theta) = Q_{\rm trans}\underbrace{\frac12(1-\cos(2\theta))}_{\displaystyle{\hat a_{\rm
 trans}}}
 - Q_{\rm diag}\,\,\underbrace{\frac12\sin(2\theta)}_{\displaystyle{\hat a_{\rm
 diag}}}\,.\label{eq:waveform}
\end{equation}

This result shows that smooth rotation of the ion trap about an
angle $\theta(t)$ is achieved by a potential that is a
time-dependent superposition of the two quadrupole fields, $\Delta
U_{\rm trans}$ and $\Delta U_{\rm diag}$. For the desired evolution,
the amplitudes $\hat a_{\rm trans}$ and $\hat a_{\rm diag}$ need to
oscillate sinusoidally at twice the rotation frequency of the trap
axis and with a relative phase shift of $\pi/2$.
Figure~\ref{fig:Waveforms}(c) illustrates the evolution of these
amplitudes during a half rotation of the trap (\emph{i.e.}\ one swap
operation). The positive offset for $\hat a_{\rm trans}$ can be
understood as an offset to render the trap circular in the plane of
rotation before adding a purely rotating two-dimensional quadrupole
field.

\begin{figure}
\begin{center}
\includegraphics[width=0.8\textwidth,keepaspectratio]{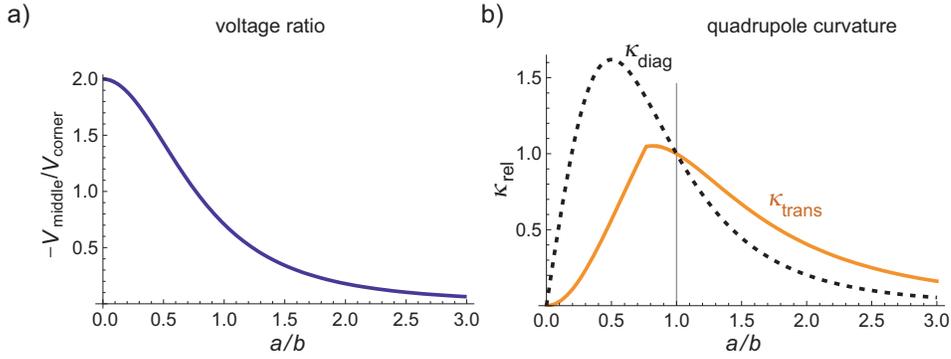}
\parbox{150mm}{\caption
{There is a certain voltage ratio $-V_{\rm middle}/V_{\rm corner}$
required between the
 (negative) middle and (positive) outer electrodes to create the transverse
 2D-quadrupole. Graph (a) shows this ratio as a
 function of the outer electrodes' aspect ratio $a/b$ (see figures~\ref{fig:TrapModel3D}
 and \ref{fig:Waveforms}). (b) Strength of the transverse (solid line) and diagonal (dotted line)
 quadrupole field for fixed maximum voltage on the electrodes. The value is normalised to
 the strength of the transverse quadrupole field at unit aspect ratio.
 At exactly unit aspect ratio both quadrupole fields have
 the same strength for a given maximum voltage.}
 \label{fig:QuadrupoleStrength}}
\end{center}
\end{figure}

One may now determine an optimum layout of electrodes to generate
the two quadrupole fields required for the trap rotation. For a
linear segmented trap a set of six electrodes is considered, as
sketched in figure~\ref{fig:TrapModel3D}, that is supplied with two
different voltage configurations in order to produce the transverse
(a) and the diagonal (b) quadrupole field, respectively. To estimate
the resulting potential the electrodes are approximated by six
(four) point charges
that are proportional to the applied voltage.
The effect of the electrodes' spatial extent, the leads and the shielding by the radio-frequency
electrodes is neglected. For these simplifying assumptions the
curvatures of the DC potential can be calculated analytically, and
the resulting fields are illustrated in
figure~\ref{fig:Waveforms}(a) and (b) respectively. The geometry is
characterised by the aspect ratio of the outermost electrodes' spacing, $a/b$ .

In order to create the transverse quadrupole field
(figure~\ref{fig:TrapModel3D}(a) and \ref{fig:Waveforms}(a)\,), the
four corner electrodes are positively charged, while the inner
electrodes are provided with negative voltage. When the ratio
$V_{\rm middle}/V_{\rm corner}$ of inner to outer electrode voltages
is adapted to the electrode aspect ratio as depicted in
figure~\ref{fig:QuadrupoleStrength}(a) ($V_{\rm middle}/V_{\rm
corner}=-2(1+(a/b)^2)^{-\frac 3 2})$, the resulting field in the
centre will be a two-dimensional quadrupole field as required (the
calculation of this is given in the appendix). The physical limitation for
the fields created is typically given by a maximum voltage that can
be applied to the DC electrodes. For aspect ratios $a/b
> 0.766$ the outer electrodes have higher voltages and will limit
the quadrupole strength, otherwise the limitation will come from the
inner electrodes. Figure~\ref{fig:QuadrupoleStrength}(b) shows the
maximum quadrupole strength, \emph{i.e.} the curvature, $\kappa$, as
given by this limitation. The strength has been normalized to the
value at unit aspect ratio, which is very close to the maximum
strength.

For the creation of the diagonally oriented quadrupole, only four
electrodes with alternating voltages, as depicted in
figure~\ref{fig:Waveforms}(b), are required. Symmetry of positive
and negative charges implies that the resulting quadrupole field
always has the desired two-dimensional character, irrespective of
the electrodes' aspect ratio. The corresponding quadrupole strength
is indicated as a dotted line in
figure~\ref{fig:QuadrupoleStrength}(b). Interestingly, at unit
aspect ratio the diagonal quadrupole has the same strength as the
transverse one, this configuration thus represents an ideal choice
for the geometry of the corner electrodes, for the transverse
quadrupole as well as for the electrodes for the diagonal
quadrupole. In a three-dimensional geometry with gap size $b$
(\emph{c.f.} figure~\ref{fig:TrapModel3D}) this configuration could
be implemented by three adjacent pairs of electrodes each having a
width of approximately half the gap size.

In light of these findings one may now interpret the experimentally
determined electrode voltages used in the realisation of the
one-point turn (see figure~\ref{fig:ElectrodeVoltages}(c)\,). Since
only three intermediate points have been determined, the curves
represent a coarse sampling of the sine and the cosine functions
that are expected (see figure~\ref{fig:Waveforms}(c)\,). The voltage
applied to the diagonal electrodes (DE) is responsible for the
diagonal quadrupole field and samples a sine function. The positive
sign accounts for the fact that the trap is rotating clockwise in
this case. The small sinusoidal amplitude of the $y$-balance (YB)
corrects for an offset between ion position and the true quadrupole
centre. An evolution similar to the cosine function is visible on
the endcap electrodes (EC) (positive sign) and the middle electrode
(ME) (negative sign), as expected. The remaining voltage excursions
compensate for residual electric fields. In particular, the long
centre electrode prevents motion perpendicular to the trap surface,
which arises from the pushing and pulling effects of endcap and
middle electrodes, since the ions are not located in the electrode
plane. Overall, the heuristically developed voltage sequence to
achieve the one-point turn in a surface trap is qualitatively
similar to the voltage requirements calculated for a non-planar
trap.

\section{Conclusion and Outlook}
\label{sec:Outlook}

In conclusion, we have demonstrated that an ion string can be
reliably rotated in a linear, segmented surface trap. Theory shows
that such rotations are possible with constant trap frequencies
throughout the duration of the turn, and that very low induced
heating rates are possible. The understanding of the potentials
involved suggests that this scheme also applies to linear segmented
traps in which the ions are located on the electrodes' symmetry
axis.\footnote{Comment added in proof: Based on this understanding,
we have successfully implemented a one-point turn in a
gold-on-ceramic two-layer trap. The trap was built as part of the EU
project MICROTRAP and is similar in design to that used in
\cite{Schulz:08}.} When this rotation is combined with the splitting
and merging of an ion string, the ions may be arranged into any
specific order without the need for additional electrodes. The
possibility of sorting ions in these traps may have an important
impact on several existing ion trapping experiments with segmented
traps. These are most notably experiments in which sympathetic
cooling of ions is applied, or for which other deterministic
rearrangement of ions is required.

\ack

We gratefully acknowledge the support of
the Austrian Science Fund (FWF),
the EU network SCALA, and the EU STREP project MICROTRAP,
IARPA,
and of the Institut f\"{u}r Quanteninformationsverarbeitung.

\appendix

\section{Comments on the point-charge model}

The point-charge model is based on the radial Coulomb potentials
originating from the six (or four) charges indicated in figure~\ref{fig:TrapModel3D}(a) (or (b)\,).
The model presented assumes charged
spheres with a radius much smaller than the distance between the
spheres. In this limit, the capacitance between different spheres is
negligible as compared to the sphere's self-capacitance, and hence
the voltages required to charge the spheres are directly
proportional to the desired charges. The charge ratio required for
the creation of a 2D-quadrupole in the $x$-$y$-plane can be obtained
by the condition that the resulting electric potential has a
vanishing curvature along the $z$-direction. For a charge $q$
located at $(x,y,0)$ this curvature, taken at the origin, turns out
to be $-q\,K/(x^2+y^2)^{3/2}$ (with $K=1/4\pi\epsilon_{\rm 0}$). If the
four corner electrodes are each charged with $Q_{\rm corner}$, they
produce a potential curvature of
\begin{equation}
 \kappa_{\rm z,C}=-4\frac{8\,Q_{\rm corner}K}{(a^2+b^2)^{3/2}}
\end{equation}
while the middle electrodes charged with $Q_{\rm middle}$ generate
the curvature
\begin{equation}
 \kappa_{\rm z,M}=-2\frac{8\,Q_{\rm middle}K}{b^3}\,.
\end{equation}
In order to make the total curvature,
$\kappa_{\rm z}=\kappa_{\rm z,C}+\kappa_{\rm z,M}$, vanish, the charge ratio of
middle to corner electrodes thus needs to be chosen as
\begin{equation}
 Q_{\rm middle}/Q_{\rm
 center}=-\frac{4b^3}{2(a^2+b^2)^{3/2}}=-\frac{2}{(1+(a/b)^2)^{3/2}}\,.
\end{equation}

Together with the proportionality between voltage
and charge in the limit of small spheres this yields the result
stated above.


\section*{References}

\end{document}